\documentclass[aps,pre,twocolumn,twoside, amsmath,amssymb,aps,bbm,floatfix,superscriptaddress]{revtex4-1}

\usepackage{xcolor}
\usepackage{graphicx}
\usepackage{subfigure}
\usepackage{dcolumn}
\usepackage{bm}
\usepackage{framed}
\usepackage{hyperref}
\usepackage{dsfont}
\usepackage[normalem]{ulem}
\usepackage{nicefrac}
\usepackage{MnSymbol}

\usepackage{tikz}

\usepackage{nicefrac}
\usepackage{braket}

\newcommand{\be}{\begin{equation}}
\newcommand{\ee}{\end{equation}}
\newcommand{\bea}{\begin{eqnarray}}
\newcommand{\eea}{\end{eqnarray}}
\newcommand{\la}{\left\langle}
\newcommand{\ra}{\right\rangle}

\newcommand{\er}[1]{\textcolor{black}{#1}}

\newcommand{\el}[1]{\textcolor{black}{#1}}
\newcommand{\pf}[1]{\textcolor{black}{#1}}

\begin{document}

\title{Efficiency at maximum power of  a Carnot quantum information engine}

\author{Paul Fadler}
 \affiliation{Department of Physics, Friedrich-Alexander-Universit\"at Erlangen-N\"urnberg, D-91058 Erlangen, Germany}
\author{Alexander Friedenberger}
 \affiliation{Department of Physics, Friedrich-Alexander-Universit\"at Erlangen-N\"urnberg, D-91058 Erlangen, Germany}
\author{Eric Lutz}
\affiliation{Institute for Theoretical Physics I, University of Stuttgart, D-70550 Stuttgart, Germany}

\begin{abstract}
 Optimizing the performance of  thermal machines is an essential task of  thermodynamics. \er{We here consider} the optimization  of information engines that convert information about the state of a system into   work. We \er{concretely introduce} a generalized finite-time Carnot cycle for a quantum information engine and  optimize its power output in the regime of low dissipation. We derive a general formula for its efficiency at maximum power valid for arbitrary working media. We further  investigate  the optimal performance of a qubit information engine subjected to  weak energy measurements.
\end{abstract}

\maketitle

Heat engines convert thermal energy into mechanical work by  running cyclicly between two heat baths at different temperatures.  They have been widely used to generate motion, from ancient steam engines to modern internal combustion motors \cite{cen01}. Information engines, on the other hand,  extract energy from a single heat bath by  processing information, for instance, via cyclic measurement and feedback operations \cite{cao09,sag10,abr11,hor11,bau12,sag12,esp12,man13,hor13,um15,par16,yam16,hor19}. They thus exploit information gained about the state of a system to produce useful work \cite{sei12,sag12a}. Such machines may be regarded as interacting with one heat reservoir and one information reservoir which  only exchanges entropy, but no energy, with the device \cite{def13,bar14,bar14a}. Information engines are possible owing to a fundamental connection between information and thermodynamics, as exemplified by Maxwell's   celebrated demon \cite{mar09,par15,lut15}. Successful information-to-work conversion has been reported in a growing number  of classical experiments  \cite{toy10,rol14,kos14,kos14a,kos15,vid16,chi17,rib19,pan18,adm18,pan20}.

At low enough temperatures,   typical nonclassical effects, such as coherent superposition of states and measurement back-action that randomly perturbs the state of a system, come into play \cite{jac14}. They deeply affect the work extraction mechanism and impact the performance of measurement controlled quantum machines  \cite{llo97,qua06,jac09,kim11,str13,bra15,elo17,elo18,sea20}. In this context, quantum measurements,  in either their  strong (projective) or weak (nonprojective) forms \cite{jac14}, may be considered    as an unconventional thermodynamic resource
 \cite{llo97,qua06,jac09,kim11,str13,bra15,elo17,elo18,sea20}.  Experimental investigations of the thermodynamic properties of a quantum Maxwell's demon,  based on  quantum measurement and feedback control of a qubit system, have recently  been performed using nuclear magnetic resonance \cite{cam16} as well as superconducting  \cite{cot17,mas18,nag18} and cavity quantum electrodynamical  \cite{naj20} setups.

Two central performance measures of heat engines  are efficiency, defined as the ratio of work output and heat input, and power that characterizes the work-output rate \cite{cen01}. The efficiency of any heat engine coupled to thermal baths is bounded from above by the Carnot efficiency, $\eta_\text{C} = 1- T_\text{c}/T_\text{h}$, where $T_\text{c,h}$ are the respective temperatures of the cold and hot heat reservoirs  \cite{cen01}. This value is usually only reachable in the ideal reversible limit, which corresponds to vanishing power.  However, real thermal machines operate in finite time with finite power, and far from reversible conditions. Their efficiency is hence reduced by irreversible losses \cite{and85,and11}. Optimizing the cyclic operation of heat engines is therefore crucial. A  practical figure of merit is the efficiency at maximum power which has been extensively studied  for classical \cite{che94,bro05,sch07,esp09,esp10,pro16} and quantum \cite{gev92,lin03,wan12,aba12,ros14} heat engines. A general  example of such an efficiency  at maximum power is  the  Curzon-Ahlborn formula, $\eta_\text{CA} = 1- \sqrt{T_\text{c}/T_\text{h}}$, which bears a striking resemblance to the Carnot expression, except for the square root \cite{cur75}. The Curzon-Ahlborn efficiency appears to be universal for finite-time Carnot machines that operate under conditions of low, symmetric dissipation \cite{esp09}. While information engines also run in finite time and with finite power,  no generic expression for their efficiency at maximum power is currently known,
owing to the difficulty to properly optimize them \cite{um15,par16,yam16}.

We here introduce a generalized  Carnot cycle for a quantum information engine by replacing the cold heat bath  of a finite-time quantum Carnot heat engine by an information reservoir. This cycle is fully reversible in the infinite-time limit. We optimize its power output  and derive a general formula for the efficiency at maximum power for arbitrary working media within the framework of  nonequilibrium thermodynamics in the weak dissipation regime. We obtain a Curzon-Ahlborn-like expression where the optimal cold coupling time is replaced by a new dissipation time that characterizes irreversible losses. We further illustrate our findings with the example of a qubit information engine, and obtain a microscopic expression of its efficiency at maximum power.
\begin{figure*}[t]
	\centering
	\begin{tikzpicture}
	\node (a) [label={[label distance=-.8 cm]145: \textbf{a)}}] at (0,0) {\includegraphics[width=0.48\textwidth]{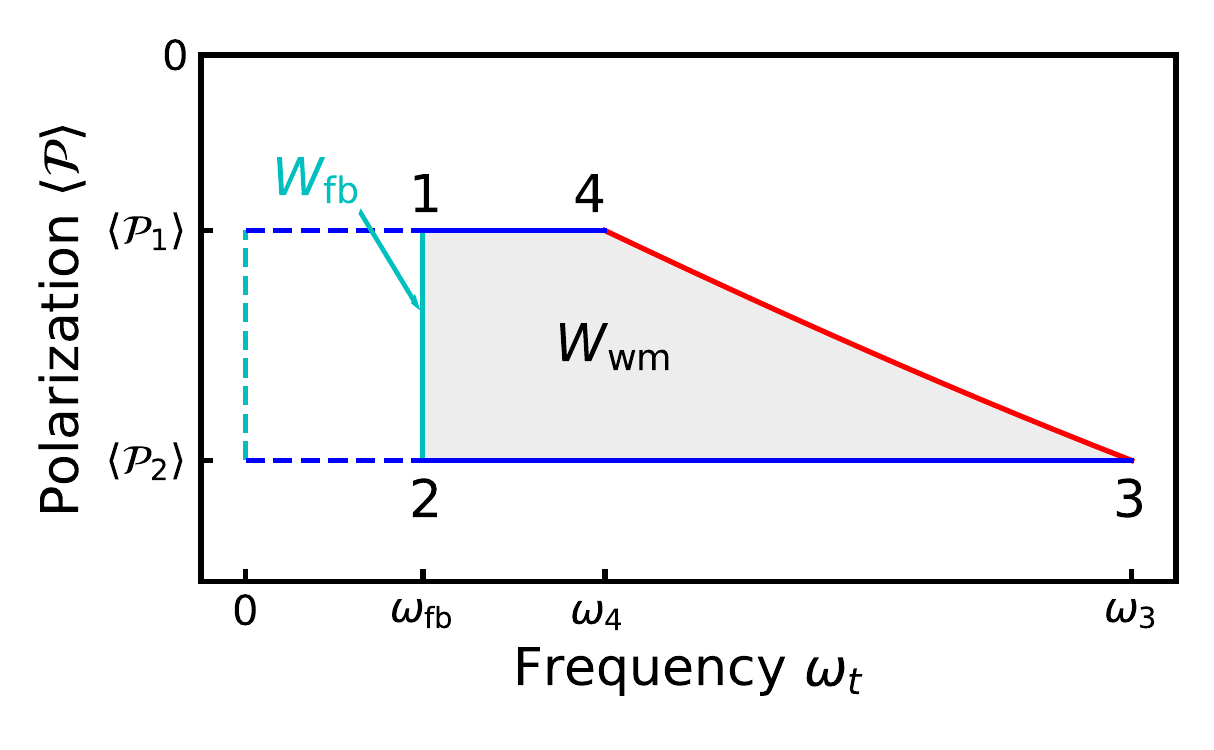}};	\node (a) [label={[label distance=-.8 cm]145: \textbf{b)}}] at (8.9,0) {\includegraphics[width=0.48\textwidth]{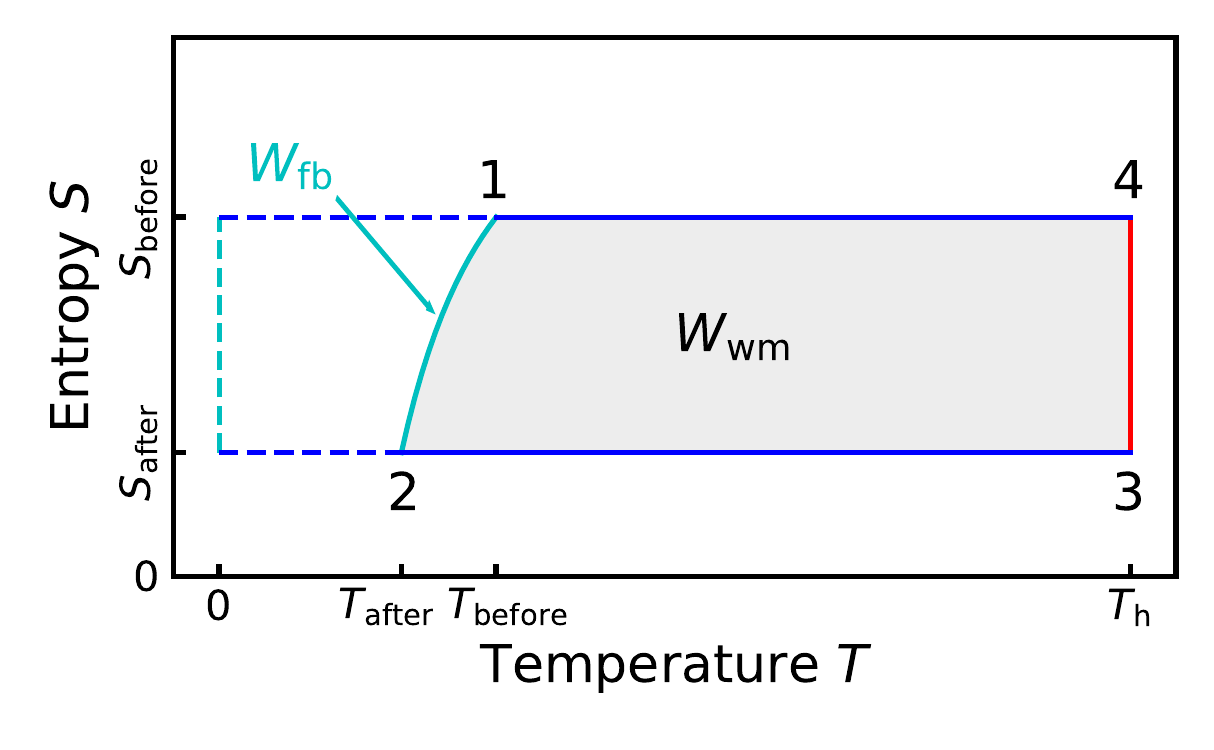}};
	\node (a) [label={[label distance=.7 cm]170: \textbf{c)}}] at (5.1,-4.7) {\includegraphics[width=0.85\textwidth]{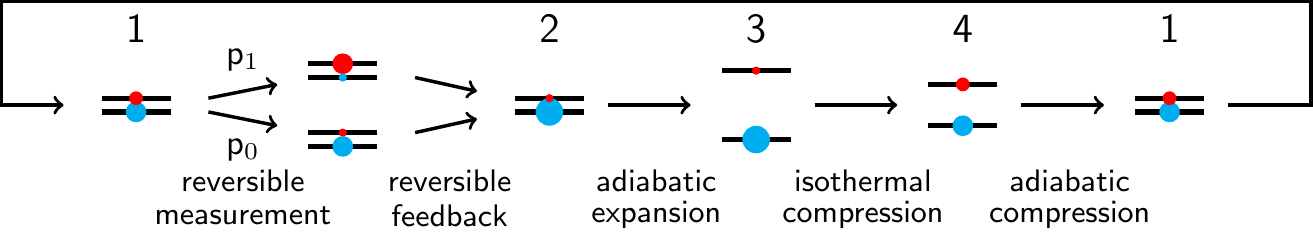}};
	\end{tikzpicture}
	\caption{Generalized finite-time Carnot cycle for the quantum information engine. a) Polarization-frequency diagram for an arbitrary working medium with Hamiltonian $H_t=\omega_t {\cal P}$. The cycle consists of one isochore during which a reversible measurement-plus-feedback protocol is implemented (1-2),   one adiabatic expansion (2-3), one isothermal compression (3-2),  and one adiabatic compression (4-1). The work  $\la W_{\mathrm{\er{wm}}} \ra$  produced by \er{the working medium during one cycle} is given by the enclosed area and the reversible feedback work $\la W_{\mathrm{fb}} \ra$ is extracted during step (1-2). The total work done is equal to the sum $\la W \ra =  \la W_{\mathrm{\er{wm}}} \ra+\la W_{\mathrm{fb}} \ra $. b) Entropy-temperature diagram of the same cycle. It reduces to a Carnot cycle for vanishing feedback frequency, $\omega_\text{fb}=0$, (dashed lines). c) Explicit  realization of the four steps of the cycle for a qubit information engine. The blue (red) dot represents the occupation probability of the ground (excited) state of the two-level system. The two outcomes of the reversible generalized  energy measurement with Kraus operators (7) occur with respective probabilities $(p_0,p_1)$.}\label{fig:1}
\end{figure*}

\textit{Reversible information engine cycle.} The reversible Carnot cycle describes the most efficient heat engine, and is thus of fundamental importance. It consists of two adiabatic   and of two isothermal (expansion and compression) branches  \cite{cen01}. Its realization  requires two heat baths: a hot bath from which heat is absorbed during the hot isotherm and a cold bath which takes on heat during the cold isotherm. Finite-time quantum Carnot cycles have been theoretically studied in Refs.~\cite{qua07,abe11,dan20,abi20,den21}. The first experimental implementation of a classical finite-time Carnot engine has been presented in Ref.~\cite{mar15}. We here construct a finite-time  generalization of the Carnot cycle for a quantum information engine by substituting  the cold heat bath (and the corresponding isotherm) by an information bath that involves   measurement  and subsequent outcome-dependent feedback (Fig.~1).

An important feature  of this  information cycle is that it is thermodynamically reversible for infinitely long cycle durations, like its thermal conterpart.  In other words, each branch, including measurement and feedback, {does} not dissipate any irreversible entropy  in that limit.  We concretely impose the following three conditions on the  engine cycle: (a) both measurement and feedback control are reversible, (b) the  cycle is independent of the measurement outcome, meaning that measurement and feedback operation always lead to the same state, irrespective of the measurement result, and (c) the state $\rho_\text{after}$ after measurement and feedback is a thermal state at  temperature $T_\text{after}$  with the same Hamiltonian $H$ as that of the state $\rho_\text{before}$ before the measurement.

We measure the state of the working medium of the information engine with a generalized measurement described by a set of positive operators $\{M_i\}$ that satisfy $\sum_i M_i^\dagger M_i =I$.
 The state after  a measurement is $\rho_i = M_i\rho_\text{before}M_i^\dagger/p_i$ with probability $p_i=  \text{Tr}[M_i\rho_\text{before}M_i^\dagger]$ \cite{jac14}. We denote by $S_i = -k \text{ Tr}[\rho_i \ln \rho_i$]  the entropy and by $E_i= \text{Tr}[\rho_i H]$ the    energy of that state ($k$ is the Boltzmann constant). \er{Such a generalized measurement  usually leads to a classical mixtures of states, implying that entropy is irreversibly produced during the process, $S(\rho_\text{{meas}}) > S(\rho_\text{before})$, where $\rho_\text{{meas}}= \sum_i p_i \rho_i$ is the density operator   averaged over all the measurement outcomes, unless  $[M_i,\rho_\text{before}]= 0$ {\cite{jac09}}. In order to make the measurement thermodynamically reversible, $S(\rho_\text{{meas}}) = S(\rho_\text{before})$, we accordingly require that the operators $M_i$ commute with the state of the system before the measurement, $[M_i, \rho_\text{before}] = 0$}.
Since the latter state is diagonal in the energy basis after the adiabatic compression branch, \el{the operators $M_i$ describe a nonprojective  measurement of the energy of the working fluid}.
We next apply reversible feedback control {\cite{hor11}} to transform  each state $\rho_i$ into the thermal state $\rho_\text{after}$. To that end, depending on the measurement outcome, we \er{reversibly} reorder the populations of $\rho_i$ so that they decrease monotonically with increasing energy, \er{while keeping the entropies $S_i$ constant}. We further shift the energy levels in order to obtain, after completion of the feedback operation,  the same Hamilton operator as that of the initial state $\rho_\text{before}$.  The explicit measurement-plus-feedback protocol for the case of a  two-level system is detailed below.

The average entropy change provided by the measurement is  $\la \Delta S \ra=   \sum_i p_i S_i - S_\text{before}\leq 0$, where $S_\text{before}$ is the entropy of  state $\rho_\text{before}$ before the measurement \cite{jac14}. Noting that after feedback control, $\rho_i =  \rho_\text{after}$ and, therefore, $S_i =  S_\text{after}$ for all measurement outcomes $i$, we simply have $\la \Delta S \ra=   S_\text{after}- S_\text{before}= \Delta S$. The average work extracted by the \er{reversibly operating} feedback controller is  additionally $\langle W_{\mathrm{fb}} \rangle= \sum_i p_i (E_i -  E_{\mathrm{after}})$, \er{since {the individual entropies $S_i$ remain} constant during the feedback process. Furthermore, since $[M_i,\rho_\text{before}]=0$, and hence $\sum_i p_i E_i = E_{\mathrm{before}}$, we  have  $\langle W_{\mathrm{fb}}  \rangle = E_{\mathrm{before}} - E_{\mathrm{after}}$.}

Let us now evaluate the  work associated with the engine cycle shown in Fig.~1. \er{For that purpose, it is useful to distinguish, on the one hand, the measurement and feedback part (step (1-2) in Fig.~1), as discussed above, and, on the other hand, the engine cycle seen from the standpoint of the working medium (steps (1-4) in Fig.~1) \cite{com1}}. During adiabatic expansion and compression, the system is isolated from the bath. \er{In order to make these steps reversible and avoid quantum friction \cite{kos02,kos03,zam14}, the Hamiltonian is chosen to  commute
with itself at all times, $[H_t , H_{t'} ] = 0$, as in the standard quantum Carnot cycle \cite{qua07,abe11,dan20,abi20,den21}}. As a result, nonadiabatic transitions do not occur for all
    driving times while work is performed. For concreteness, and without loss of generality, we consider a Hamilton operator of the scaling form $H_t= \omega_t {\cal P}$, with time-dependent frequency $\omega_t$ \cite{den21}. From the point of view of the working medium, the cycle then consists of four branches (Fig.~1): (1-2)  one isochore at constant frequency $\omega_\text{fb}$, (2-3) one adiabat with frequency variation from $\omega_\text{fb}$ to $\omega_3$, (3-4) one isotherm with frequency change from $\omega_3$ to $\omega_4$ at constant \el{bath} temperature $T_\text{h}$, and (4-1) one adiabat   with frequency decrease from $\omega_4$ to $\omega_\text{fb}$. The average produced work $\langle W_{\mathrm{\er{wm}}}\rangle$ is simply given by the area enclosed by the cycle. According to the first law \er{applied to the working medium}, we have $\langle W_{\mathrm{\er{wm}}}\rangle= \langle Q_\text{h}\rangle+\langle Q_\text{c}\rangle$, where $\langle Q_\text{h,c}\rangle$ are the respective heat contributions from the isotherm and the isochore. In the long-time limit, the heat absorbed from the hot reservoir may be written in leading order (low dissipation regime) as $Q_\text{h}= T_\text{h} ( \Delta S - {\Sigma}/{\tau_\text{h}})$, where $\Sigma$ is a coefficient that characterizes  the entropy production  during time $\tau_\text{h}$  along the isotherm \cite{esp10}. Moreover, the heat \er{exchanged by the working medium during the cold isochore can be evaluated by purely thermodynamic means (without involving the measurement and feedback aspect) \cite{gev92,lin03,wan12}}. It is given by $\langle Q_\text{c}\rangle= \omega_\text{fb} \Delta \langle {\cal P}\rangle=E_{\mathrm{after}} - E_{\mathrm{before}} $.

\er{The total work $\la W\ra$  done during  the complete information engine cycle is  the sum of the work extracted by the feedback controller, $\la W_{\mathrm{fb}} \ra$, and the work produced by the working medium, $\la W_{\mathrm{wm}} \ra$}. We hence obtain
 \begin{equation}
 \label{1}
\la \er{W} \ra = \la W_{\mathrm{fb}} \ra + \la W_{\mathrm{\er{wm}}} \ra= T_\text{h} \left( \Delta S - \frac{\Sigma}{\tau_\text{h}} \right).
\end{equation}
    We note that  $\la Q_\text{c}\ra$ and {$\la W_{\mathrm{fb}}\ra$} exactly cancel.  In other words, the information reservoir  only exchanges entropy but no energy with the system. We are now in the position to investigate the phenomenological finite-time performance of the generalized Carnot information engine.

\textit{Efficiency at maximum power.}  The efficiency at which information is converted into work in the cyclic quantum information engine is defined as \cite{llo97,qua06,jac09,kim11,str13,bra15,elo17,elo18,sea20}
\begin{equation}
\label{2}
\eta = \frac{\la \er{W}\ra}{T_\text{h} \Delta S} = 1 - \frac{\Sigma}{\Delta S \tau_\text{h}},
\end{equation}
where we have used Eq.~\eqref{1}. Unit efficiency ($\eta_\text{max}=1$) is achieved for $\tau_\text{h}\rightarrow \infty$, when the cycle is  reversible. In this regime, information about the state of the system, gained through the measurement, is fully converted into  work by the  cyclic engine. For finite-time operation, the efficiency is reduced ($\eta <1$)  owing to dissipative processes associated with irreversible entropy production.

The  power of the information engine further reads \cite{cen01}
\begin{equation}
  \label{4}
P = \frac{\la \er{W}\ra}{\tau_\text{h} + \tau_{\mathrm{fb}}}=\frac{T_\text{h} \left( \Delta S - \frac{\Sigma}{\tau_\text{h}} \right)}{\tau_\text{h} + \tau_{\mathrm{fb}}},
\end{equation}
where $\tau_{\mathrm{fb}}$ denotes the time of the measurement and feedback protocol. The time spent along the  two adiabats can be set to zero since they  are reversible irrespective of their duration \cite{gev92,lin03}. By contrast, the feedback time $\tau_{\mathrm{fb}}$ is determined by the measurement-feedback process and we take it to be fixed \er{\cite{com}}.
Setting the derivative of the power $P$ with respect to $\tau_\text{h}$ to zero, we find the    optimal coupling time to the hot heat reservoir
\begin{equation}
  \label{4}
\tau^*_\text{h} = \frac{\Sigma}{\Delta S} \left( 1 + \sqrt{1 + \frac{\Delta S}{\Sigma}\tau_{\mathrm{fb}}} \right).
\end{equation}
The corresponding efficiency at maximum power $\eta^* $ of the quantum information engine then follows as
\begin{equation}
\label{5}
\eta^* = 1 - \frac{1}{1+\sqrt{1 + \tau_{\mathrm{fb}}/\tau_\text{h}^{\er{\oast}}}} = 1- \frac{\tau_\text{h}^{\er{\oast}}}{\tau_\text{h}^*},
\end{equation}
where we have used Eq.~\eqref{4} and introduced the typical dissipation time $\tau_\text{h}^{\er{\oast}}=\Sigma/\Delta S$ associated with irreversible losses along the hot isotherm: $\tau_\text{h}^{\er{\oast}}$ is small (resp. large) when the entropy production is small (resp. large). Expression \eqref{5} is reminiscent of the Curzon-Ahlborn formula \cite{cur75}, which can be written in terms of the optimal cold and hot coupling times, $\tau_\text{c}^*$ and $\tau_\text{h}^*$,  as $\eta_\text{CA}= 1-\tau_\text{c}^*/\tau_\text{h}^*$ \cite{gev92}. The optimal time of the cold isotherm $\tau_\text{c}^*$ is here simply replaced by the new dissipation time $\tau_\text{h}^{\er{\oast}}$. We moreover observe \el{from Eq.~\eqref{5}} that in general $\eta_\text{max}/2<\eta^*<\eta_\text{max}=1$, the lower (upper) bound being reached when the  feedback time is much smaller (larger) than the dissipation time $\tau_\text{fb}\ll \tau_\text{h}^{\er{\oast}}$ ($\tau_\text{fb}\gg \tau_\text{h}^{\er{\oast}}$).

With the help of the above expressions, the maximum power $P^{*}$ may furthermore be written as,
\begin{equation}
P^{*} = \frac{\eta^{*}T_\text{h} \Delta S}{\tau_{\mathrm{h}}^{*} + \tau_{\mathrm{fb}}},
\end{equation}
with the optimal produced work $\la \er{W} \ra^*= \eta^* T_\text{h} \Delta S$. These results generically hold for any working medium.

\begin{figure*}[t]
	\centering
	\begin{tikzpicture}
	\node (a) [label={[label distance=-.10 cm]145: \textbf{a)}}] at (0,0) {\includegraphics[width=0.48\textwidth]{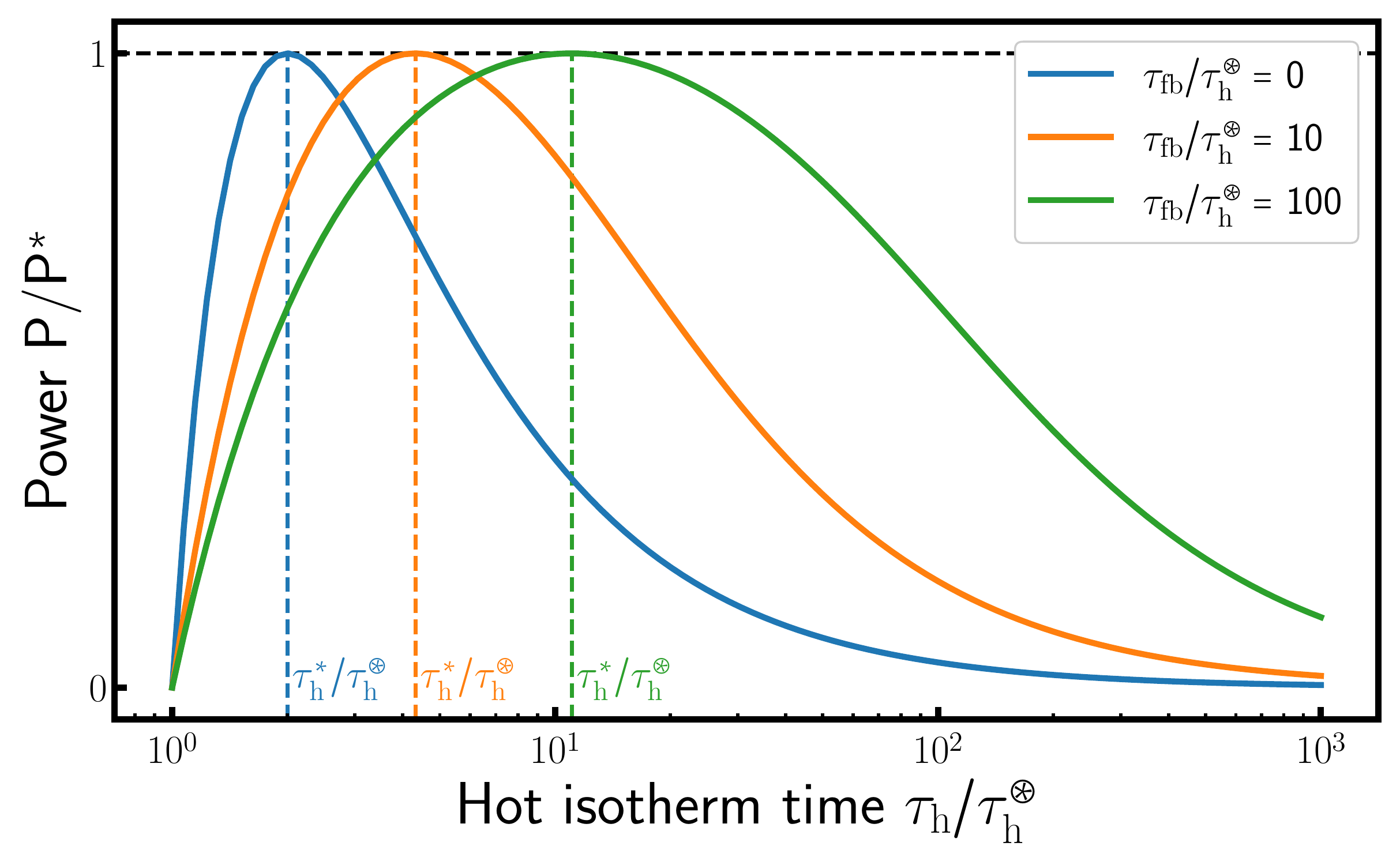}};	\node (a) [label={[label distance=-.1 cm]145: \textbf{b)}}] at (8.9,0) {\includegraphics[width=0.5\textwidth]{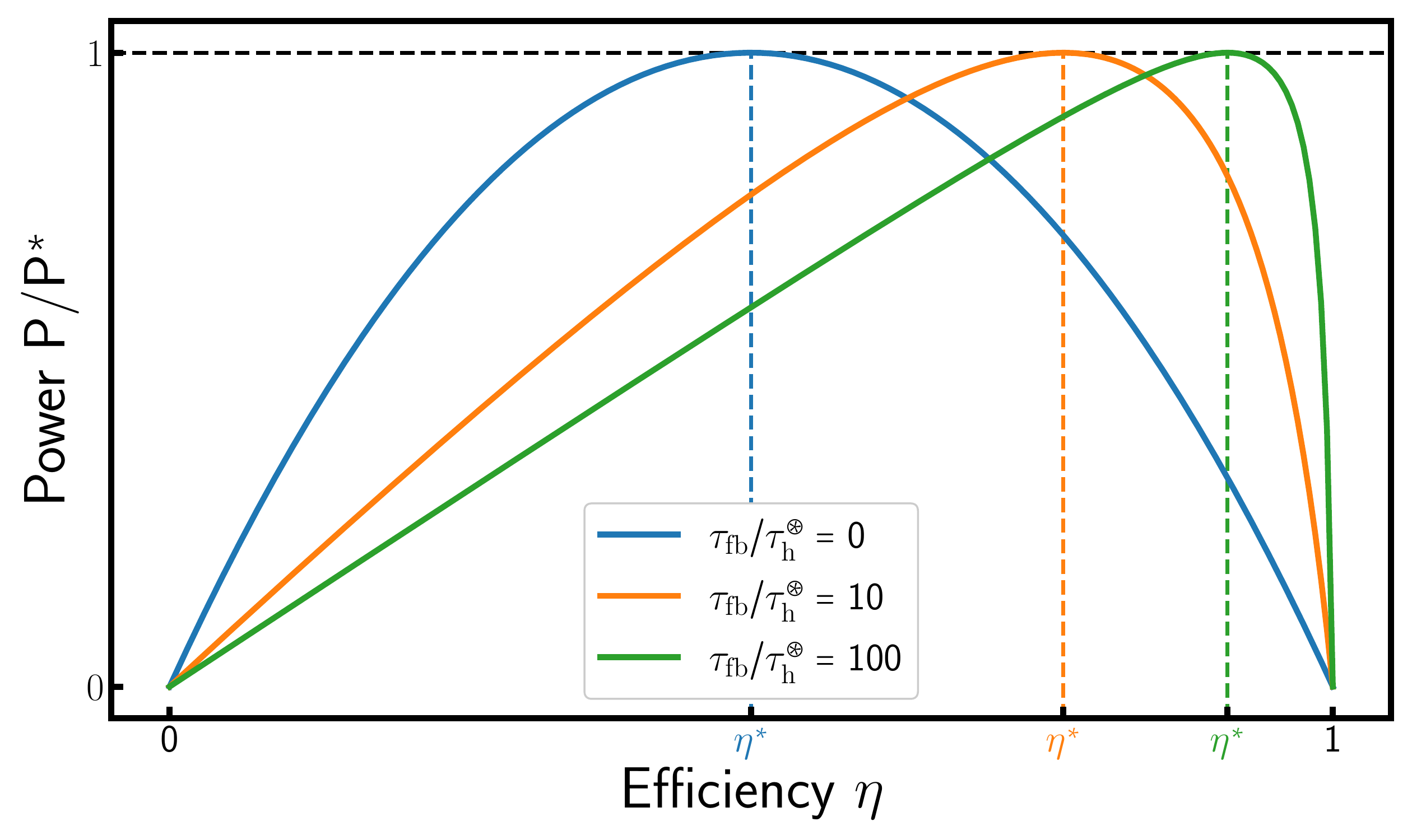}};
	\end{tikzpicture}
	\caption{Optimal performance of the quantum information engine. a) Reduced power $P/P^*$, Eq.~(3), as a function of the duration of the hot isotherm $\tau_\text{h}$, Eq.~(4), for different values of  the feedback time $\tau_\text{fb}$ (both in units of the dissipation time $\tau_\text{h}^{\er{\oast}}$). Maximum power $P^*$ is reached at the optimal time $\tau_\text{h}^*$. b) Power versus efficiency curves, for the same parameters, that exhibit the characteristic shape of an endoreversible engine. The general inequality $\eta_\text{max}/2<\eta^*<\eta_\text{max}=1$ is verified.} \label{fig:2}
\end{figure*}

\textit{Qubit information engine.} We proceed by illustrating our findings with the case of a  spin-1/2 information engine with Hamilton operator $H_t= \omega_t \sigma_z/2 = \omega_t {\cal P}$, where $\sigma_z$ is the usual Pauli operator and ${\cal P}=\sigma_z/2$ is the polarization. The knowledge  of the precise quantum dynamics of this system allows for the microscopic evaluation of the efficiency at maximum power of the information engine.

 We begin by specifying the measurement-feedback protocol of the generalized finite-time Carnot cycle (Fig.~1). In order to satisfy the  conditions (a)-(c) stated above (measurement and feedback should be reversible, all measurement results should be mapped onto the thermal state $\rho_\text{after}$ with the same Hamilton operator as  $\rho_\text{before}$), we construct a generalized  quantum measurement such that the first  measurement outcome ($i=0$) is   $\rho_\text{after}$ (that is, $\rho_0= \rho_\text{after}$ with energy $E_0= E_\text{after}$) and the second  measurement outcome ($i=1$) is equal to \el{its spin-flipped counterpart (that is, $\rho_1= \sigma_x\rho_\text{after}\sigma_x$ with energy $E_1= -E_\text{after}$)}. The corresponding measurement operators are explicitly given by (Supplemental Material \cite{sup})
\begin{eqnarray}
\label{6}
M_0 &=& \sqrt{\frac{1-e^{(\beta_\text{b}+ \beta_\text{a}) \omega_\text{fb}}}{1-e^{2\beta_\text{a}\omega_\text{fb}}}}\ket{1} \bra{1} + \sqrt{\frac{1-e^{-(\beta_\text{b}+ \beta_\text{a}) \omega_\text{fb}}}{1-e^{-2\beta_\text{a} \omega_\text{fb}}}} \ket{0} \bra{0} \nonumber \\
M_1 &=& \sqrt{\frac{1-e^{(\beta_\text{b}- \beta_\text{a}) \omega_\text{fb}}}{1-e^{-2\beta_\text{a} \omega_\text{fb}}}}\ket{1} \bra{1} +\sqrt{\frac{1-e^{-(\beta_\text{b}- \beta_\text{a}) \omega_\text{fb}}}{1-e^{2\beta_\text{a}\omega_\text{fb}}}} \ket{0} \bra{0}\nonumber \\
\end{eqnarray}
where $\beta_\text{b}=\beta_\text{before}$ and  $\beta_\text{a}=\beta_\text{after}$ are the respective inverse temperatures of the states $\rho_\text{before}$ and $\rho_\text{after}$. The kets $\ket{0}$ and $\ket{1}$ denote the (ground and excited) energy eigenstates of the qubit. The Kraus operators \eqref{6} describe a \el{nonprojective} energy measurement   of the spin-1/2 \el{(it  becomes weak in the high-temperature limit)}.

We next apply outcome-dependent feedback control to transform all the measurement results $(i= 0,1)$ into the same state $\rho_\text{after}$. For  outcome $0$, we apply the identity $I$, since $\rho_0= \rho_\text{after}$ by construction; we hence trivially have  $H_0= H$. For outcome $1$, we unitarily rearrange the states with the transformation $H_1 = -H + (E_1 - E_{\mathrm{after}}) I$, which leaves the energy of the state unchanged, $\text{Tr}[\rho_1 H_1] = \text{Tr}[\rho_1H]$. We finally shift the energy level to  obtain the Hamiltonian of the state $\rho_\text{before}$. In doing so, we extract the feedback work $\langle W_{\mathrm{fb}} \rangle  = E_{\mathrm{before}} - E_{\mathrm{after}}$.

The interaction of the two-level system with the hot heat bath may be microscopically described with the help of a usual quantum master equation of the form  \cite{gev92,lin03}
\begin{eqnarray}
\label{8}
  \dot{\cal P}_t &= &  \gamma_+ \left( \sigma_-[{\cal P}_t,\sigma_+] + [\sigma_-,{\cal P}_t]\sigma_+ \right)\nonumber\\
  &+ &\gamma_- \left( \sigma_+[{\cal P}_t,\sigma_-] + [\sigma_+,{\cal P}_t]\sigma_-  \right) +\frac{\partial {\cal P}_t}{\partial t},
\end{eqnarray}
for the polarization ${\cal P}_t$ in the Heisenberg picture and the operators $\sigma_\pm = \sigma_x \pm i \sigma_y$. Assuming that the damping coefficients satisfy the detailed-balance condition $\gamma_-/\gamma_+= \exp(\beta_\text{h}\omega_t)$, by choosing, for instance, the  concrete parametrization $\gamma_+= a \exp(q\beta_\text{h} \omega_t)$ and $\gamma_-=a \exp((1+q)\beta_\text{h}\omega_t)$ (with $a>0$ and $0>q>-1$ constant parameters), Eq.~\eqref{8} can be rewritten as \cite{gev92,lin03}
\begin{equation}
\dot{ \la  {\cal P}_t\ra} = -a e^{q\beta_\text{h}\omega_t}[2(1+e^{\beta_\text{h}\omega_t})\la  {\cal P}_t\ra+ (e^{\beta_\text{h}\omega_t}-1)].
\end{equation}
The parameter $a$  characterizes the magnitude of the damping coefficients and, thus, the rate of change of the average polarization.
Solving  the above equation for time \cite{gev92,lin03}, the duration of the isotherm in the high-temperature limit ($\beta_\text{h} \omega_{3,4}\ll1$) is found to read \cite{sup}
\begin{equation}
  \label{9}
\tau_\text{h} = \frac{\ln \left({\omega_3}/{\omega_4} \right)}{4a \left( 1 - {\beta_\text{h}}/{\beta'} \right)},
\end{equation}
where the effective inverse temperature $\beta'$ of the qubit is determined via $\la{\cal P}_t\ra= -\tanh(\beta'\omega_t/2)/2$ \cite{gev92,lin03}. Due to the finite-time relaxation of the system, the temperature $T'$ is not necessarily equal to the bath temperature $T_\text{h}$, when thermalization is not complete; we have $\tau_\text{h}  \rightarrow \infty$ when $T'\rightarrow T_\text{h}$ (or $a\rightarrow 0$). Noting further that the work $\la \er{W} \ra = T_\text{h} ( \Delta S - {\Sigma}/\tau_\text{h})$ produced by the irreversible engine cycle with bath temperature $T_\text{h}$ is  equal to the work $T' \Delta S$ produced by a reversible cycle with effective bath temperature $T'$ \cite{gev92}, we find the dissipation time,
\begin{equation}
\label{11}
\tau_\text{h}^{\er{\oast}}= \frac{\Sigma}{\Delta S} =  \frac{\ln ( {\omega_3}/{\omega_4})}{4a}.
\end{equation}
Equation \eqref{11} is solely determined by the beginning and end frequencies $\omega_{3,4}$ of the isotherm and the bath coupling parameter $a$. We therefore obtain the microscopic expression for the efficiency at maximum power \eqref{5}:
\begin{equation}
\label{12}
    \eta^* =1- \frac{\tau_\text{h}^{\er{\oast}}}{\tau_\text{h}^*}= 1 - \frac{1}{1+\sqrt{1 + 4a\tau_{\mathrm{fb}}/\ln ( {\omega_3}/{\omega_4})}}.
\end{equation}

Figure 2a) displays the reduced power $P/P^*$ of the qubit information engine as a function of the duration of the hot isotherm $\tau_\text{h}$ for different values of the feedback time $\tau_\text{fb}$ (both in units of $\tau_\text{h}^{\er{\oast}}$). We identify a clear maximum at the optimal time $\tau_\text{h}^*$ given by Eq.~\eqref{4}. Figure 2b) moreover shows the corresponding power versus efficiency curves that are typical for an endoreversible engine \cite{che94}. Such  machines are internally reversible   and irreversible losses only occur via thermal contact with the external bath. They hence outperform fully irreversible engines and have played for this reason a central role in finite-time thermodynamics \cite{and85,and11}. We note that the general inequality $\eta_\text{max}/2<\eta^*<\eta_\text{max}=1$ is satisfied.

\textit{Conclusions.}
We have proposed a generalized finite-time Carnot cycle for a quantum information engine. Like the standard Carnot cycle for heat engines, it is thermodynamically reversible for large cycle durations. This cycle thus describes the most efficient quantum information engine with unit information efficiency. We have optimized its power output in the regime of low dissipation and derived a Curzon-Ahlborn-like formula for its efficiency at maximum power. This generic expression only depends on the optimal time of the hot isotherm and a new dissipation time associated with irreversible entropy production. The efficiency at maximum power was further shown to obey the general inequality $1/2<\eta^*<1$, independent of the microscopic details of the engine. Our results provide a theoretical basis for the optimization of information engines. We hence expect them to be important for the  study of optimal quantum machines in finite-time information thermodynamics.

\begin{acknowledgments}
  We acknowledge  financial assistance from the German Science Foundation (DFG) (under project FOR 2724) and thank
  Florian Marquardt for his support.
 \end{acknowledgments}

\onecolumngrid
\begin{appendix}
\er{\section{Alternative thermodynamic analysis}}

\er{Instead of performing the thermodynamic analysis of the finite-time quantum information engine as done in the main text by distinguishing, on the one hand, the measurement and feedback part (step (1-2) in Fig.~1), and, on the other hand, the engine cycle seen from the standpoint of the working medium (steps (1-4) in Fig.~1), we here present an alternative, but, equivalent,  description that separates the \pf{measurement in point 1 in Fig.~1} and the remaining cycle which is now considered to be operated by a general controller. The latter cycle includes the feedback part that is conditioned on the measurement outcome, as well as the two adiabats and the isotherm (steps (1-4) in Fig.~1). This derivation generalizes the one discussed in Ref.~\cite{bra15} for a different quantum information cycle \pf{through the usage of incomplete measurements}.}

\er{We begin by evaluating the respective  changes of energy and entropy associated with (i) the measurement and with (ii) the remaining cycle. We have
\begin{align}
  \Delta E^{\mathrm{meas}}(m) &= E[\rho_{m}] - E[\rho_1]\\
  \Delta S^{\mathrm{meas}}_{\mathrm{sys}}(m) &= S_{\mathrm{sys}}[\rho_{m}] - S_{\mathrm{sys}}[\rho_1]
\end{align}
for  the measurement with outcome $m$, and
\begin{align}
  \Delta E^{\mathrm{cyc}}(m) &= E[\rho_1] - E[\rho_{m}]\\
  \Delta S^{\mathrm{cyc}}_{\mathrm{sys}}(m) &= S_{\mathrm{sys}}[\rho_1] - S_{\mathrm{sys}}[\rho_{m}]
\end{align}
for the cycle implemented by the controller.
The total entropy production during the \pf{cycle} is the sum of the entropy change of the system and of the bath
\begin{equation}
  \Delta S_{\mathrm{tot}}^{\mathrm{cyc}}(m) = \Delta S^{\mathrm{cyc}}_{\mathrm{sys}}(m) + \Delta S_{\mathrm{bath}} \geq 0\pf{,}
\end{equation}
since the total entropy production is \pf{non-negative. We} accordingly obtain
\begin{equation}
  Q_\text{h}= Q(m) = -T\pf{\Delta S_{\mathrm{bath}}}\leq T \Delta S^{\mathrm{cyc}}_{\mathrm{sys}}(m).
\end{equation}
The first law applied to the complete control operation then reads
\begin{align}
  \Delta E^{\mathrm{cyc}}(m) &= Q(m) - W(m) =Q_\text{h} - W,
  \end{align}
  or, equivalently,
  \begin{align}
  W(m) &\leq \pf{T \Delta S^{\mathrm{cyc}}_{\mathrm{sys}}(m)} - \Delta E^{\mathrm{cyc}}(m).
\end{align}
Since the measurement is reversible, that is,  $S[\rho_{m}] = S[\rho_2]$ for all $m$,
averaging $\Delta E^{\mathrm{cyc}}(m)$ yields
\begin{align}
  \left\langle \Delta E^{\mathrm{cyc}} \right\rangle
  &= \sum_{m}p_{m}\left( E[\rho_1] - E[\rho_{m}] \right)
  = E[\rho_1] - \sum_{m}p_{m}\mathrm{Tr}\left( H\rho_{m} \right)
  = E[\rho_1] - \sum_{m}p_{m}\mathrm{Tr}\left( H\frac{M_{m}\rho_1M_{m}^{\dagger}}{p_{m}} \right)\\
  &= E[\rho_1] - \mathrm{Tr}\left( H\rho_1 \sum_{m}M_{m}^{\dagger}M_{m} \right)
  = E[\rho_1] - \mathrm{Tr}\left( H\rho_1 \right) = 0,
\end{align}
where we have used $[H,M_{m}]=0$ and $\sum_{m}M_{m}^{\dagger}M_{m}=1$. Similarly, we have $ \left\langle \Delta E^{\mathrm{meas}}\right\rangle =0 $.\\
Since $\Delta S^{\mathrm{meas}}_{\mathrm{sys}}= \pf{-}\Delta S^{\mathrm{cyc}}_{\mathrm{sys}} = S_{\mathrm{sys}}[\rho_2]-S_{\mathrm{sys}}[\rho_1]$, we eventually arrive at
\begin{equation}
  \left\langle W \right\rangle \leq -T \Delta S^{\mathrm{meas}}_{\mathrm{sys}} = -T(S[\rho_2] - S[\rho_1]).
\end{equation}}

\er{In the low dissipation limit, we may be further write the entropy production in the form
\begin{align}
  \frac{\Sigma}{\tau_h}&= \Delta S^{\mathrm{cyc}}_{\mathrm{sys}}(m) + \Delta S_{\mathrm{bath}}(m),
  \end{align}
from which we find
  \begin{align}
  Q_\text{h}=- Q(m) &= -T\Delta S_{\mathrm{bath}}(m) = T\left( \Delta S^{\mathrm{cyc}}_{\mathrm{sys}}(m) - \frac{\Sigma}{\tau_h}\right)
\end{align}
Combining everything, we finally obtain (denoting $\Delta S^{\mathrm{\pf{cyc}}}_{\mathrm{sys}} = \Delta S $, as in the main text)
\begin{equation}
  \left\langle W \right\rangle = T \left( \Delta S - \frac{\Sigma}{\tau_h} \right).
\end{equation}
This is equation (1) of the main text.}\\

\section{Measurement operators for the qubit information engine}
We here explicitly derive the Kraus operators $M_i$ for the generalized quantum measurement implemented in the two-level information engine. They  have to fulfill the condition
\begin{equation}
  \label{eq:condKraus}
\rho_{\mathrm{after}} = \Phi_i[\rho_i],
\end{equation}
where the state $\rho_\text{after}$ (after measurement and feedback) is a thermal state at effective temperature $T_\text{after}$ and $\rho_i = M_i\rho_{\mathrm{before}}M_i^{\dag}/p_i$ is the state  of the system after a measurement with outcome $i=(0,1)$. For $i=0$, $\Phi_0 = {I}$ is the identity, whereas for $i=1$, $\Phi_1 = \Phi_{\mathrm{flip}}$ is the quantum bit flip channel.

Let us parametrize the thermal states before and after the measurement as
\begin{align}
  \rho_{\mathrm{before}} &= \alpha\ket{1}\bra{1} + (1-\alpha)\ket{0}\bra{0},\\
  \rho_{\mathrm{after}} &=  \beta\ket{1}\bra{1} + (1-\beta)\ket{0}\bra{0}.
\end{align}
Since the  operators $M_i$ commute with thermal states, we can also parametrize them by their diagonal entries as
\begin{align}
  M_0 &= x\ket{1}\bra{1} + y\ket{0}\bra{0},\\
  M_1 &= u\ket{1}\bra{1} + v\ket{0}\bra{0}.
\end{align}
Neglecting arbitrary phases by choosing $(x,y,u,v) \in \mathds{R}^+$, which can always be done by properly adjusting the adiabatic protocol, we can eliminate the parameters $u$ and $v$ by using
\begin{align}
 \mathds{1} =  M_0^{\dag}M_0 + M_1^{\dag}M_1 = \left( x^{*}x + u^{*}u \right)&\ket{1}\bra{1}
  + \left( y^{*}y + v^{*}v \right)\ket{0}\bra{0}.
\end{align}
We then obtain
\begin{align}
  u &= \sqrt{1 - x^2} \quad \text{and} \quad
  v = \sqrt{1 - y^2}.
\end{align}
Looking at measurement outcome $0$, we further have
\begin{align}
  \frac{M_0\rho_{\mathrm{before}}M_0^{\dag}}{\mathrm{Tr}\left(M_0\rho_{\mathrm{before}}M_0^{\dag}\right)} &= \rho_{\mathrm{after}},
  \end{align}
or, explicitly
\begin{align}
  \frac{x^2\alpha}{x^2\alpha + y^2(1-\alpha)}\ket{1}\bra{1} &+ \frac{y^2(1-\alpha)}{x^2\alpha + y^2(1-\alpha)}\ket{0}\bra{0}
  = \beta\ket{1}\bra{1} + (1-\beta)\ket{0}\bra{0}.
\end{align}
Setting the populations of the excited state to be equal, we find
\begin{equation}
  \label{eq:first}
\frac{x^2\alpha}{x^2\alpha + y^2(1-\alpha)} = \beta.
\end{equation}
The equality of the populations of the ground state is  automatically  fulfilled owing to the unit trace.
On the other hand, looking at measurement outcome $1$, we have
\begin{equation}
  \frac{M_1\rho_{\mathrm{before}}M_1^{\dag}}{\mathrm{Tr} \left( M_1\rho_{\mathrm{before}}M_1^{\dag} \right)}
  = \Phi_{\mathrm{flip}}[\rho_{\mathrm{after}}],
\end{equation}
which leads  to the equation
\begin{equation}
  \label{eq:second}
  \frac{(1-x^2)\alpha}{(1-x^2)\alpha + (1-y^2)(1-\alpha)} = 1-\beta.
\end{equation}
Solving the set of Eqs.~\eqref{eq:first} and \eqref{eq:second}, we obtain
\begin{align}
  x = \sqrt{\frac{\beta - \beta^2 - \alpha\beta}{\alpha - 2\alpha\beta}}\quad \text{and} \quad
  y = \sqrt{\frac{1 - \beta}{1-\alpha}\frac{1 - \beta - \alpha}{1 - 2\beta}},
\end{align}
as well as
\begin{align}
  u &= \sqrt{1 - \beta\frac{1 - \beta - \alpha}{\alpha - 2\alpha\beta}}\quad \text{and} \quad
  v = \sqrt{1 - \frac{1 - \beta}{1-\alpha}\frac{1 - \beta - \alpha}{1 - 2\beta}}.
\end{align}
Writing further $\alpha = \frac{1}{1+e^a}$ and $\beta = \frac{1}{1+e^b}$, with $a = \beta_\text{b}{\omega_\text{fb}}$ and $b = \beta_\text{a} {\omega_\text{fb}}$, where $\beta_\text{a}= \beta_\text{after}$ is the inverse temperature of state $\rho_\text{after}$ and $\beta_\text{b}= \beta_\text{before}$ is the inverse temperature of state $\rho_\text{before}$, we finally arrive at
\begin{align}
  x 
    &= \sqrt{\frac{e^{a+b}-1}{e^{2b} - 1}}, \quad  \quad
  y 
    = \sqrt{\frac{1-e^{-a-b}}{1 - e^{-2b}}},\quad  \quad
 u = \sqrt{\frac{1 - e^{a-b}}{1 - e^{-2b}}}\quad \text{and} \quad
 v = \sqrt{\frac{e^{-a+b} + 1}{e^{2b} - 1}}.
\end{align}

\section{Hot  isotherm time for the qubit information engine}
We next derive Eq.~(10) of the main text for the duration of the hot isotherm following Refs.~\cite{gev92,lin03}. We begin with Eq.~(9) of the main text for the average polarization ${\cal P}_t$ in the Heisenberg picture
\begin{equation}
\dot{ \la  {\cal P}_t\ra} = -a e^{q\beta\omega_t}[2(1+e^{\beta_\text{h}\omega_t})\la  {\cal P}_t\ra+ (e^{\beta_\text{h}\omega_t}-1)].
\end{equation}
Applying the  chain rule for derivatives, we have
\begin{equation}
  \dot{\la  {\cal P}_t\ra} = \frac{d\la  {\cal P}_t(\omega_t)\ra}{d\omega_t}\frac{d\omega_t}{dt}.
\end{equation}
Since the effective temperature along the isothermal branch is  constant, we obtain from $\la \mathcal{P}_t(\omega_t)\ra = -\tanh \left( \beta^\prime \omega_t/2 \right)/2$
\begin{align}
 \frac{d\la\mathcal{P}_t\ra}{d\omega_t} &= -\frac{\beta^{'}}{2[\cosh(\beta^{'}\omega_t) + 1]}.
\end{align}
Combining both equations, we obtain
\begin{align}
  \frac{d\la\mathcal{P}_t\ra}{d\omega_t} \frac{d\omega_t}{dt}
  = &-a e^{q\beta_\text{h}\omega_t}[-(1+e^{\beta_\text{h}\omega_t})\tanh\left( \beta^{'}\omega_t/2 \right) + e^{\beta_\text{h}\omega_t}-1].
 \end{align}
or, equivalently, solving for $dt$
\begin{align}
  dt &= -\frac{(d\la\mathcal{P}_t\ra/d\omega_t)}{a e^{q\beta_\text{h}\omega_t}[-(1+e^{\beta_\text{h}\omega_t})\tanh\left( \beta^{'}\omega_t/2 \right) + (e^{\beta_\text{h}\omega_t}-1)]}d\omega_t,\\
  t &= \frac{1}{2a}\int_{x_0}^{x(t)}\left[ e^{q\epsilon x}(e^{\epsilon x}-e^{x})(1+e^{-x})\right]^{-1}dx,
\end{align}
where we have introduced the new variable $x=\beta'\omega_t$ and defined $\epsilon=\beta_\text{h}/\beta'$.
Using the expansion $e^x \simeq 1+x$ in the high-temperature limit, we finally arrive at
\begin{align}
  t &= \frac{1}{2a}\int_{x_0}^{x_t} \frac{dx}{(1 + q\epsilon x)(1 + \epsilon x - (1 + x))(1+1 - x)}=\frac{\ln(\omega_3/\omega_4)}{4a(1-\beta_{\mathrm{h}}/\beta^{'})}.
\end{align}

\section{Entropy production for the qubit information engine}
We finally derive an expression for the parameter $\Sigma$ which determines  the nonequilibrium entropy production. We consider an arbitrary $N$-dimensional working fluid with Hamilton operator $\mathcal{H}_t = \omega_t \mathcal{P}= \omega_t \sum_n \lambda_n \ket{n}\bra{n}$.  We choose the global energy offset of $\mathcal{P}$ such that
\begin{align}
  \sum_n \lambda_n &= 0
\quad \text{and} \quad \sum_n \lambda_n^2 = 2\chi.
\end{align}
In the high temperature limit ($\beta\lambda_n\omega_t \ll 1$), we can express the thermal state as
\begin{align}
  \rho &= \frac{e^{-\beta\omega_t \mathcal{P}}}{\mathrm{Tr}\left[ e^{-\beta\omega_t \mathcal{P}} \right]}
  \simeq\frac{(1 - \beta\omega_t\lambda_N + (\beta\omega_t\lambda_N)^2/2)\ket{N}\bra{N}}{Z},
\end{align}
with the partition function
\begin{align}
  Z &\simeq \sum_n \left[ 1 - \beta\omega_t\lambda_n + (\beta\omega_t\lambda_n)^2/2\right]
    = N - \beta\omega_t\sum_n\lambda_n + \beta^2\omega_t^2\sum_n\lambda_n^2/2
    = N + \beta^2\omega_t^2\chi.
\end{align}
The occupation probabilities are therefore
\begin{align}
  P_n &= \bra{n}\rho\ket{n}\simeq \frac{1 - \beta\omega_t\lambda_n + \beta^2\omega_t^2\lambda_n^2/2}{N + \beta^2\omega_t^2\chi}
  = \frac{1}{N} - \frac{\beta\omega_t\lambda_n}{N} + \frac{\beta^2\omega_t^2\lambda_n^2}{2N} - \frac{\beta^2\omega_t^2\chi}{N^2}.
\end{align}
The entropy can accordingly be written as
\begin{align}
  S &= -\text{Tr}[\rho \ln \rho]= -\sum_n P_{n} \ln P_n \simeq \ln N - \frac{\beta^2\omega_t^2\chi}{N}.
\end{align}
The entropy changes for an isochoric process from $\beta_1$ to $\beta_2$ at frequency $\omega_t$ and an isothermal transformation  from $\beta_1$ to $\beta_2$ at inverse temperature $\beta$ thus read
\begin{align}
  \Delta S_{\mathrm{isochoric}} = \frac{\omega_t^2\chi}{N}\left( \beta^2_2 - \beta^2_1 \right) \quad \text{and} \quad
  \Delta S_{\mathrm{isothermal}} = \frac{\beta^2\chi}{N}\left( \omega^2_2 - \omega^2_1 \right).
\end{align}
The dissipation constant $\Sigma$ for a two-level system (with $N=2$ and $\chi=1/4$)  hence follows as
\begin{align}
 {\Sigma} &= {\Delta S} \,\frac{\ln(\omega_2/\omega_1)}{4a} = {\Delta S_\text{isothermal}}\, \frac{\ln(\omega_2/\omega_1)}{4a}
= \frac{\beta^2}{8} \left( \omega^2_2 - \omega^2_1 \right)\frac{\log(\omega_2/\omega_1)}{4a}.
\end{align}

%
\end{appendix}

\end{document}